\documentclass{emulateapj}
\usepackage{graphicx}

\slugcomment {submitted to the Monthly Notices of the Royal Astronomical Society}

\shorttitle{Optical Emission from G179.0+2.6}

\shortauthors{How et al.}

\begin{document}

\title{Optical Emission Associated with the Galactic Supernova Remnant G179.0+2.6}

\author{Thomas G.\ How\altaffilmark{1},
        Robert A.\ Fesen\altaffilmark{2}, 
        Jack M.\ M.\ Neustadt\altaffilmark{2}, 
        Christine S.\ Black\altaffilmark{2}, and
        Nicolas Outters\altaffilmark{3}
 }
\altaffiltext{1}{Curdridge Observatory, Southhampton UK }
\altaffiltext{2}{Department of Physics \& Astronomy, Dartmouth
                 College, Hanover, NH 03755 USA}
\altaffiltext{3}{Orange Observatory, Haute-Savoie France }

\begin{abstract}

Narrow passband optical images of the large Galactic supernova remnant
G179.0+2.6 reveal a faint but nearly complete emission shell dominated by
strong [\ion{O}{3}] 4959,5007 \AA \ line emission.  The remnant's optical
emission, which consists of both diffuse and filamentary features, is brightest
along its southern and northeastern limbs.  Deep H$\alpha$ images detect little
coincidence emission indicating an unusually high [\ion{O}{3}]/H$\alpha$
emission ratio for such a large and apparently old remnant.  Low-dispersion
optical spectra of several regions confirm large [\ion{O}{3}]/H$\alpha$ line
ratios with typical values around 10.  The dominance of [\ion{O}{3}] emission
for the majority of the remnant's optical filaments suggests shock velocities
above 100 km s$^{-1}$ are present throughout most of the remnant, likely reflecting a
relatively low density ambient ISM. The remnant's unusually strong [\ion{O}{3}]
emission adds to the remnant's interesting set of properties which include a
thick radio emission shell, radial polarization of its radio emission like that
typically seen in young supernova remnants, and an unusually slow-rotating
gamma-ray pulsar with a characteristic spin-down age $\simeq$ 50 kyr. 

\end{abstract}
\bigskip


\section{Introduction}

Of the nearly 300 currently confirmed Galactic supernova remnants (SNRs)
\citep{Green14}, the majority were initially identified in the radio due to
nonthermal radio emissions associated with shocked gas
\citep{Milne70,Downes71}.  Of these, only $\sim$30\% exhibit 
coincident optical emission.

Here we report the discovery of extensive optical emission associated with the
large (diameter $\sim 70^{\prime}$) shell-type Galactic SNR G179.0+2.6.  This
remnant was first detected in the radio at 1.42 and 2.695 GHz by
\citet{Fuerst86} who noted a triple polarized source near its center, which was
later determined to be a background radio galaxy with bipolar jets
\citep{Fuerst89}. 

The remnant's radio emission consists of an unusually thick shell (see Fig.\ 1)
compared to similarly large and old SNRs.  It has an estimated diameter
$\simeq$ 70 pc assuming a distance estimate of $\sim$3.5 kpc based on its
surface brightness and the $\Sigma-D$ relation of \citet{Milne79}. Such a size
implies an old remnant with an age $>$10$^{4}$ yr \citep{Fuerst89}. A
radio source near its center is a gamma-ray pulsar (PSR J0554+3107) with an
unusually long spin period of 0.465 s with a characteristic age of $\simeq$52 kyr
\citep{Pletsch13}.

While the remnant's overall radio emission spectral index $\alpha = -0.45 \pm
0.11$ \citep{Gao11} is unremarkable for an old SNR, its radial projected
magnetic field is unexpected and more typical of young SNRs
\citep{Fuerst86,Fuerst89}.  \citet{Gao11}, confirmed the radial arrangement of
the remnant's magnetic field and suggested the Galactic magnetic field in the
remnant's area might be oriented along the line of sight as a possible
explanation, although they noted this was unlikely.

\begin{figure}
\centerline{\includegraphics[angle=0,width=9.6cm]{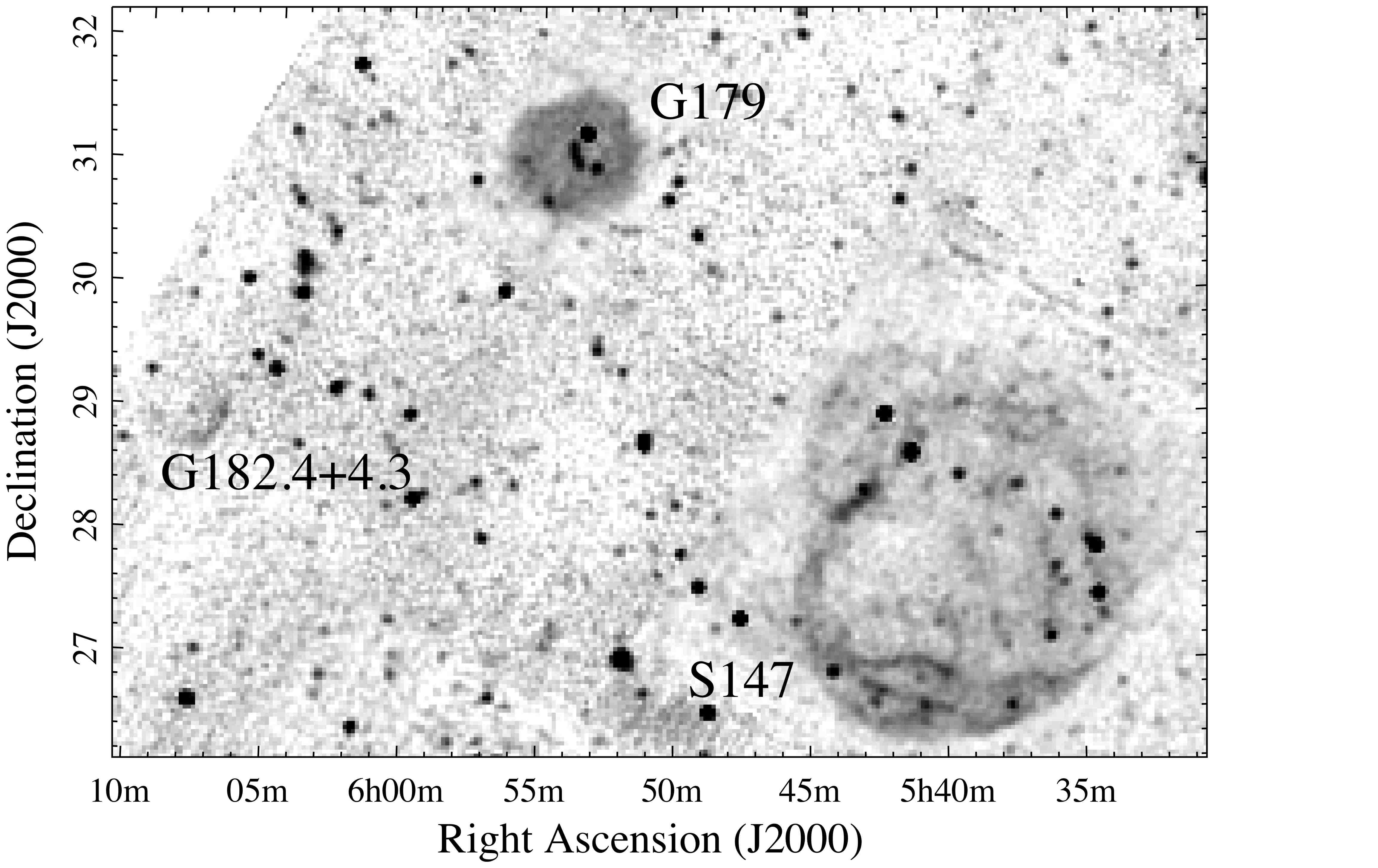}}
\caption{2695 MHz radio image of G179.0+2.6 and its surroundings from \citet{Fuerst90}, 
highlighting its unusually thick radio emission shell 
relative to the nearby SNRs S147 and G182.4+4.3. }
\label{radio_image}
\end{figure}

There has been no previously reported optical emission associated with the
G179.0+2.6 SNR (hereafter referred to as simply G179). No
coincident emission is visible on the Palomar Sky Surveys (DSS1 and DSS2), or
the emission-line survey of the Galactic plane of \citet{Parker79}.  However, a
partial shell of faint H$\alpha$ emission can be seen at the remnant's position
in the Virginia Tech Spectral Line Survey (VTSS) of the Galactic Plane
\citep{Dennison98,Fink03} and small patches of emission filaments especially
along the remnant's northeastern limb are visible in the Isaac Newton Telescope
(INT) Photometric H$\alpha$ Survey (IPHAS) of the Northern Galactic Plane
\citep{Drew05,GS08}.

\begin{figure*}[t]
\begin{center}
\leavevmode
\includegraphics[scale=0.80]{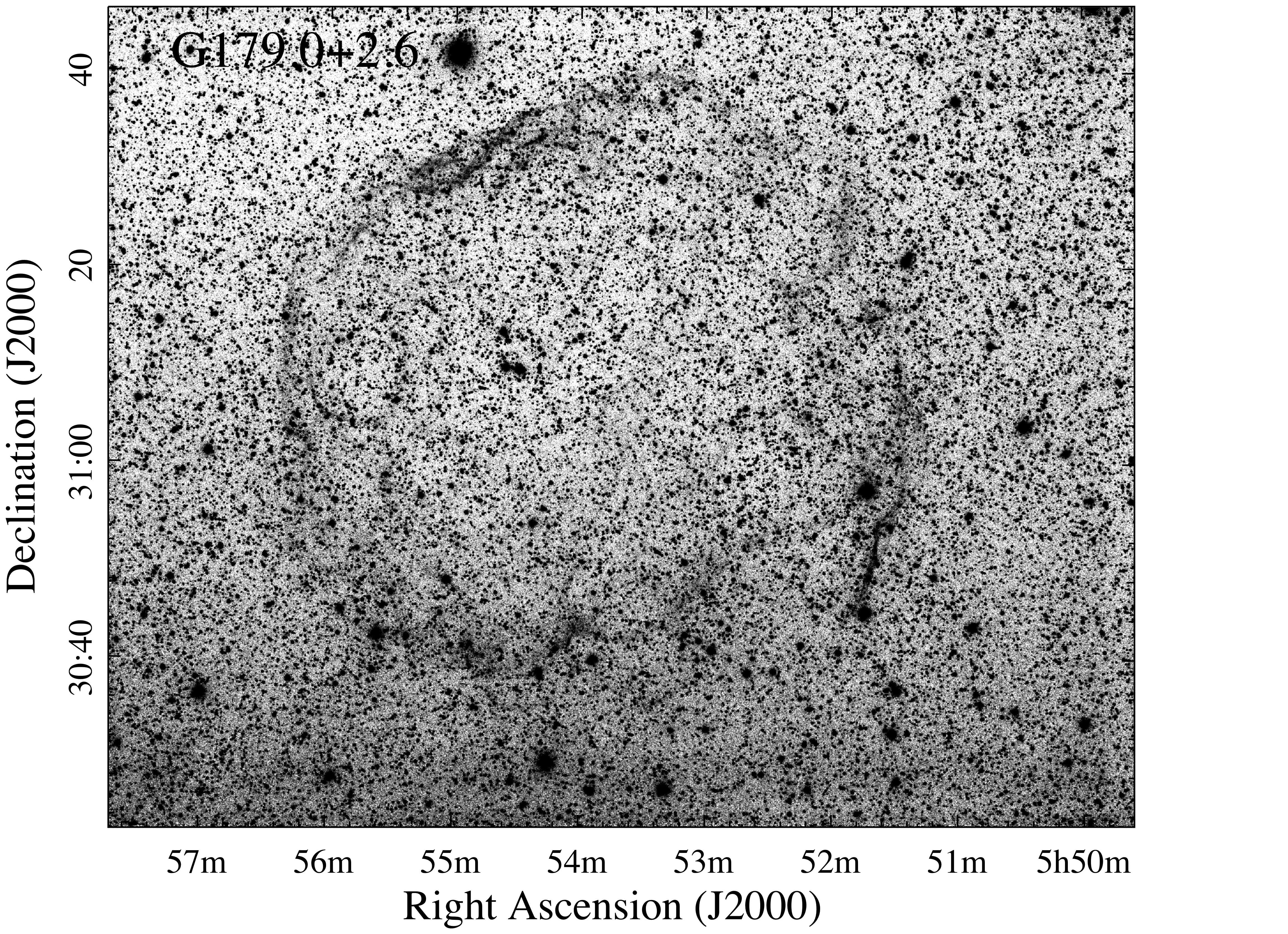}
\hspace{1em}
\includegraphics[scale=0.80]{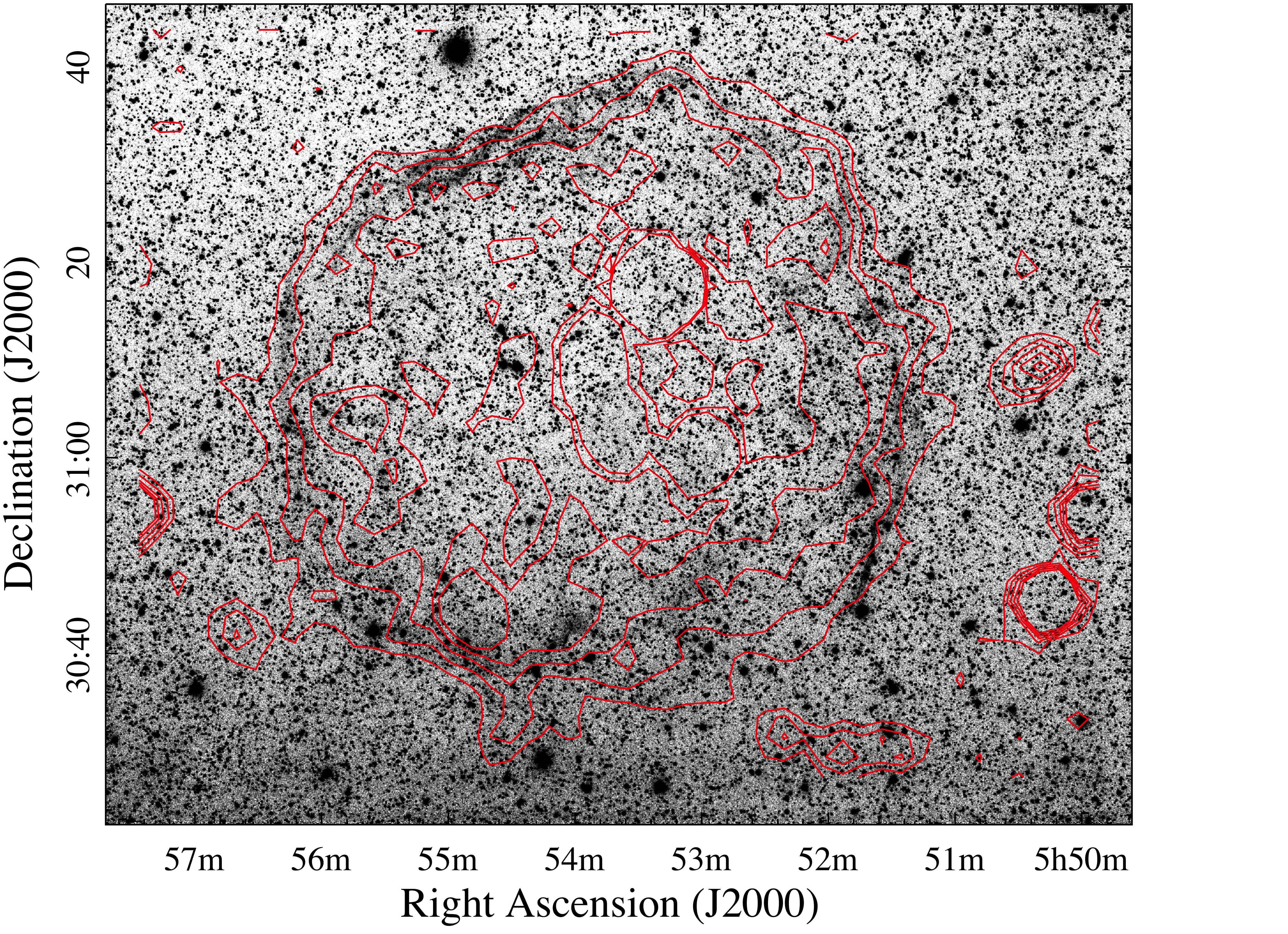}
\caption{ \textit{Top}: A 9.2 hour Curdridge Observatory [O III]  5007 \AA \ image of 
G179.0+2.6 showing a nearly complete shell of emission.
\textit{Bottom}: Same image with overlay of the 100 m Effelsberg 2695 MHz brightness temperature 
contours \citep{Fuerst90}.
}
\end{center}
\label{Toms_image}
\end{figure*}

\begin{figure*}[t]
   \begin{center}
   \leavevmode
   \includegraphics[scale=0.85]{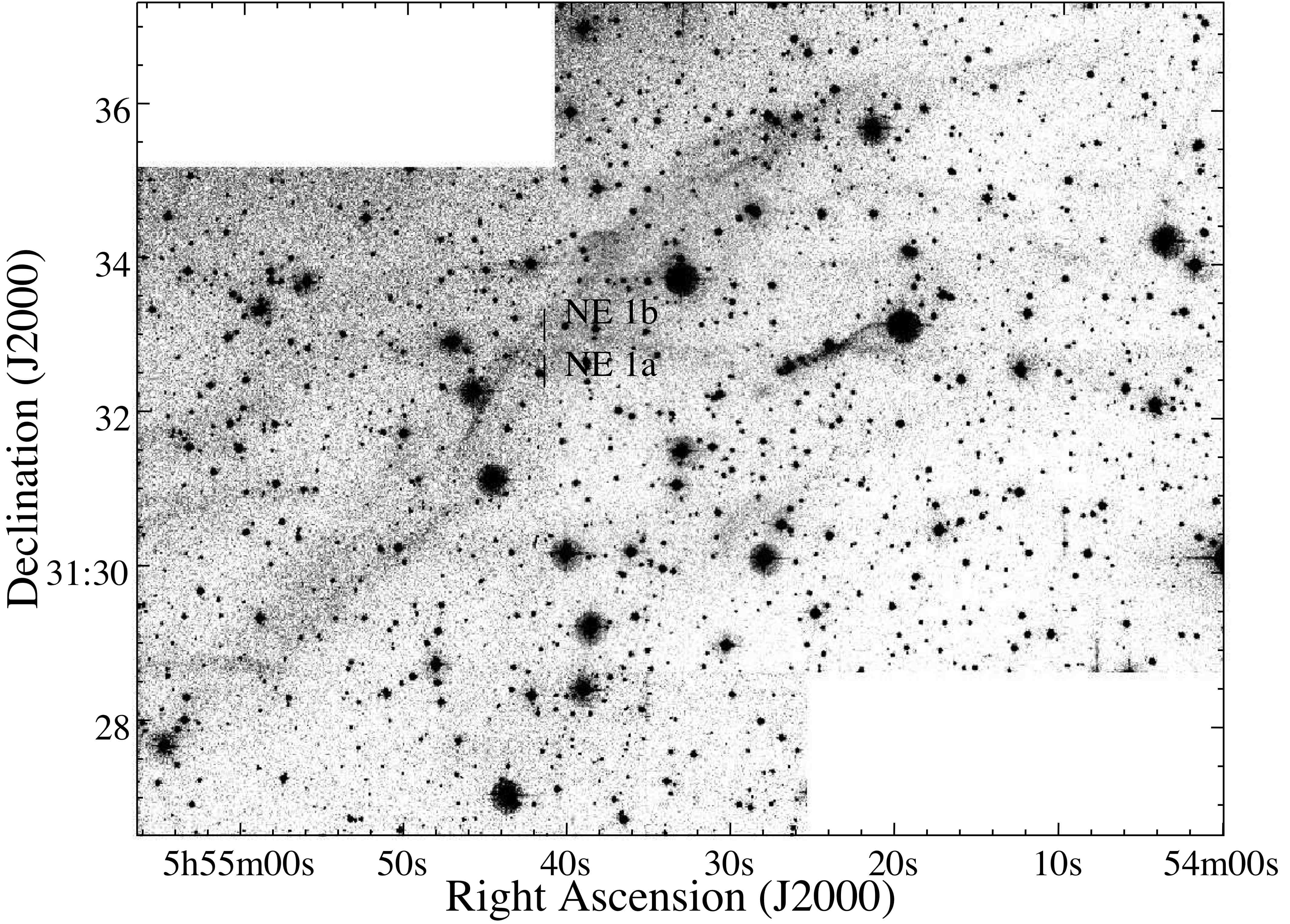}
   \hspace{1em}
   \includegraphics[scale=0.85]{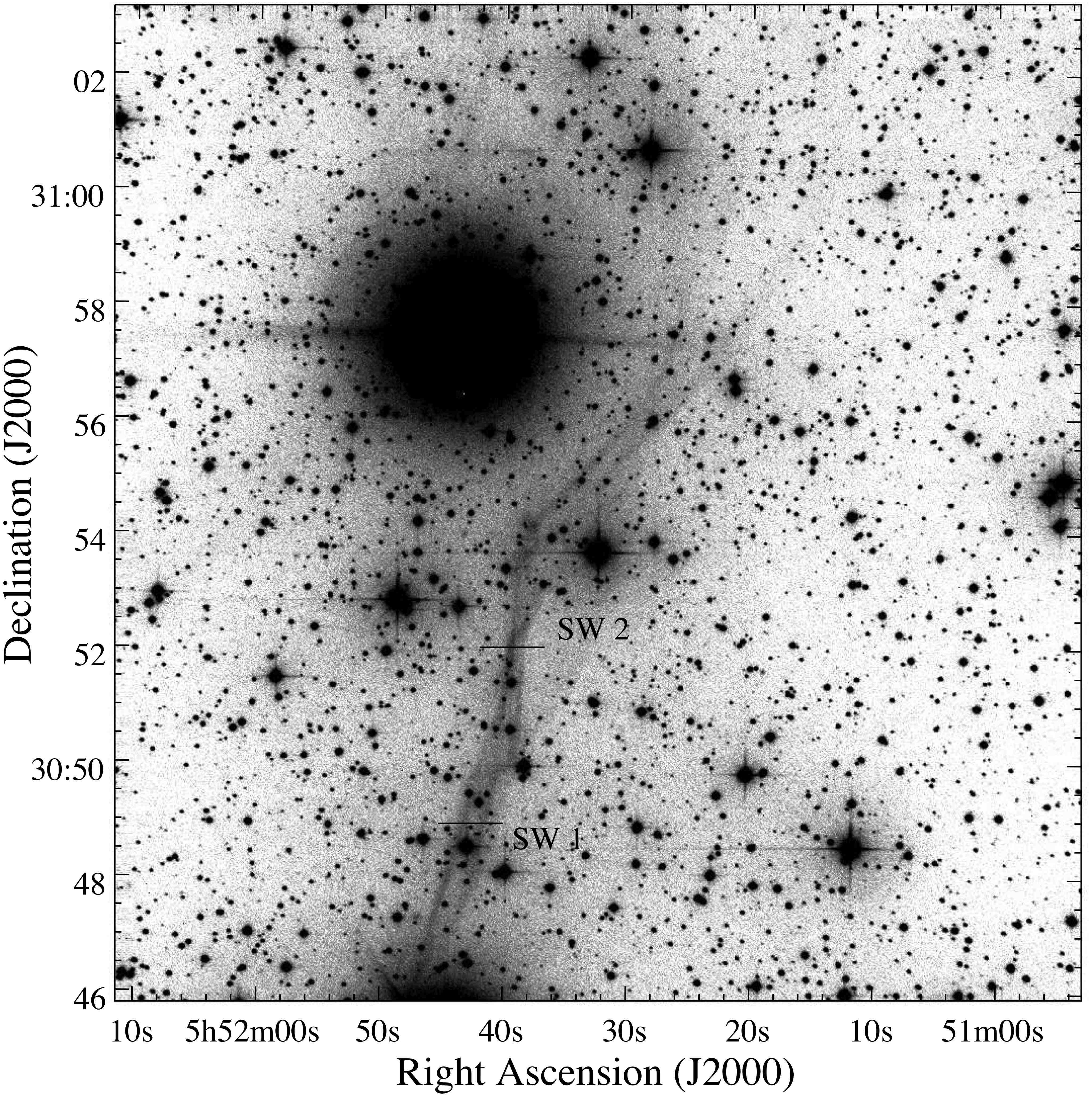}
\caption{MDM 1.3m [O III] images of northeastern region (top) and southwestern regions (bottom) of G179.0+2.6
with slit positions indicated where low-dispersion spectra were obtained. }
\end{center}
\label{MDM_images}
\end{figure*}


One of us (TH) obtained deep H$\alpha$ and [\ion{O}{3}] filter images of
G179 in a search of associated optical emission. These images revealed a
surprisingly complete [\ion{O}{3}] emission shell coincident with the remnant's
radio shell. Despite similarly deep H$\alpha$ images, little
coincident H$\alpha$ emission was detected. 

This discovery was followed-up by additional [\ion{O}{3}] and
H$\alpha$ images plus low-dispersion optical spectra of a few representative
filaments. Optical images and spectra of G179 are described in $\S$2
with results presented in $\S$3  and discussed in $\S$4.

\section{Observations}

Discovery of wide-spread optical emission from the G179 supernova remnant was
made in November 2016 using a 90 mm refractor
at the Curdridge Observatory in Southampton, UK.  Images  were taken using a
Astrodon 30 \AA \ FWHM filter and a Atik 490 camera employing a Sony ICX814 CCD
which was binned 2 $\times$ 2 yielding $1690 \times 1352$ pixels.  This system
provided a FOV of $64^{\prime} \times 86^{\prime}$ with an image scale of
3$\farcs72$ pixel$^{-1}$.  

Combining 22 dithered 1500 s exposures for a total exposure time of 9.2 hr,
faint [\ion{O}{3}] emission was detected coincident with the remnant's radio
emission.  Similarly long 30 \AA \ FWHM H$\alpha$ filter exposures showed
little coincident emission.

Follow-up images were obtained using a 0.4 m telescope at Orange
Observatory in eastern France.  Narrow passband H$\alpha$ and [\ion{O}{3}]
images (FWHM = 30 \AA) were taken using a Astrodon and a Moravian G4 camera
with a KAF-16803 CCD with $4096 \times 4096$ pixels.  This system
yielded a FOV of  $83^{\prime} \times 83^{\prime}$ with an image scale of
1$\farcs22$ pixel$^{-1}$.  

A series of 35 dithered 600 s H$\alpha$ images and 66 dithered 600 s
[\ion{O}{3}] images were obtained between December 2016 and March 2017. These 
were then combined into a 5.8 hr H$\alpha$ exposure and a 11 hr [\ion{O}{3}]
exposure of the remnant. 

\begin{figure*}[t]
\centering
\includegraphics[scale=0.65]{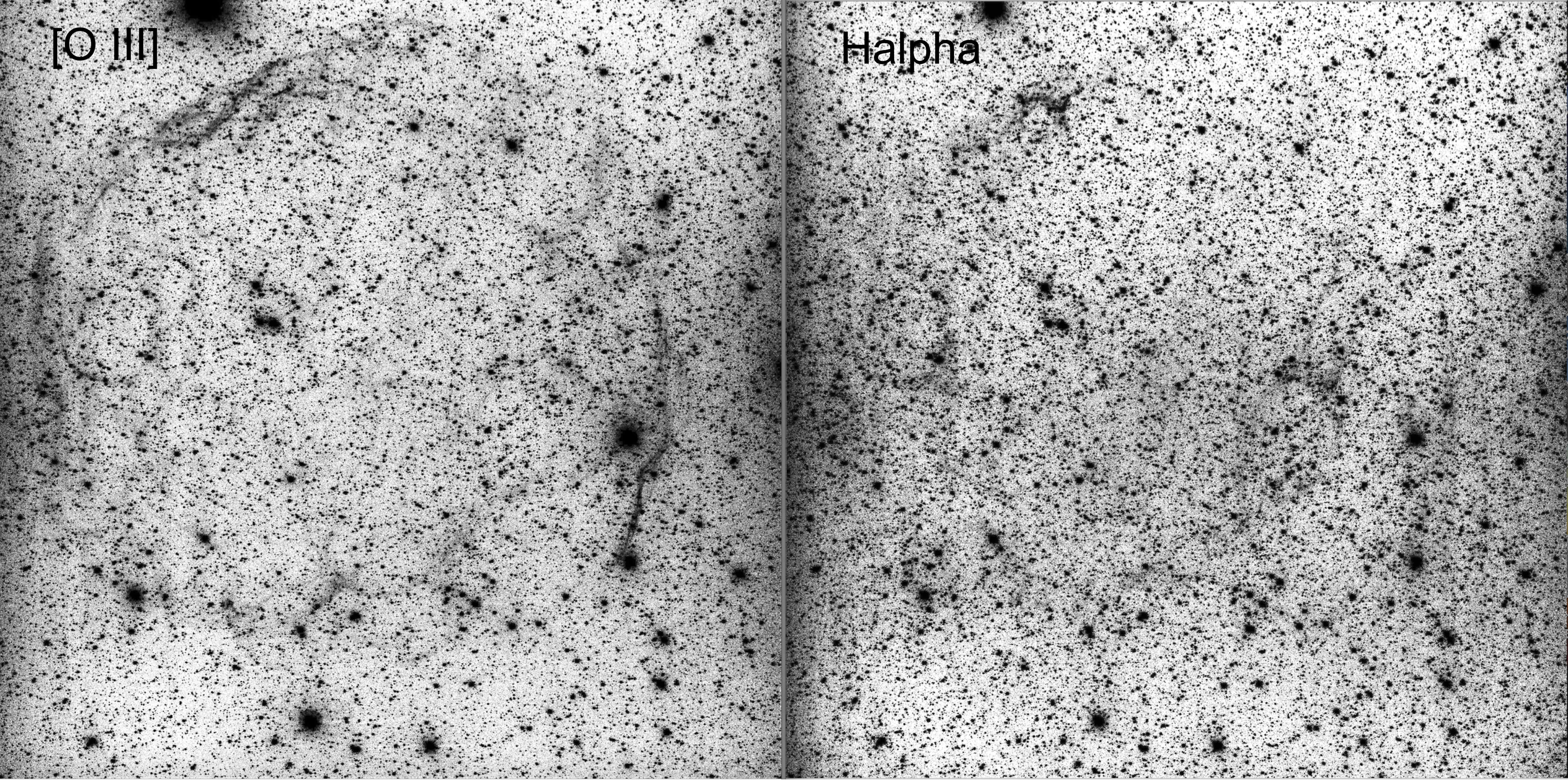}
\includegraphics[scale=0.95]{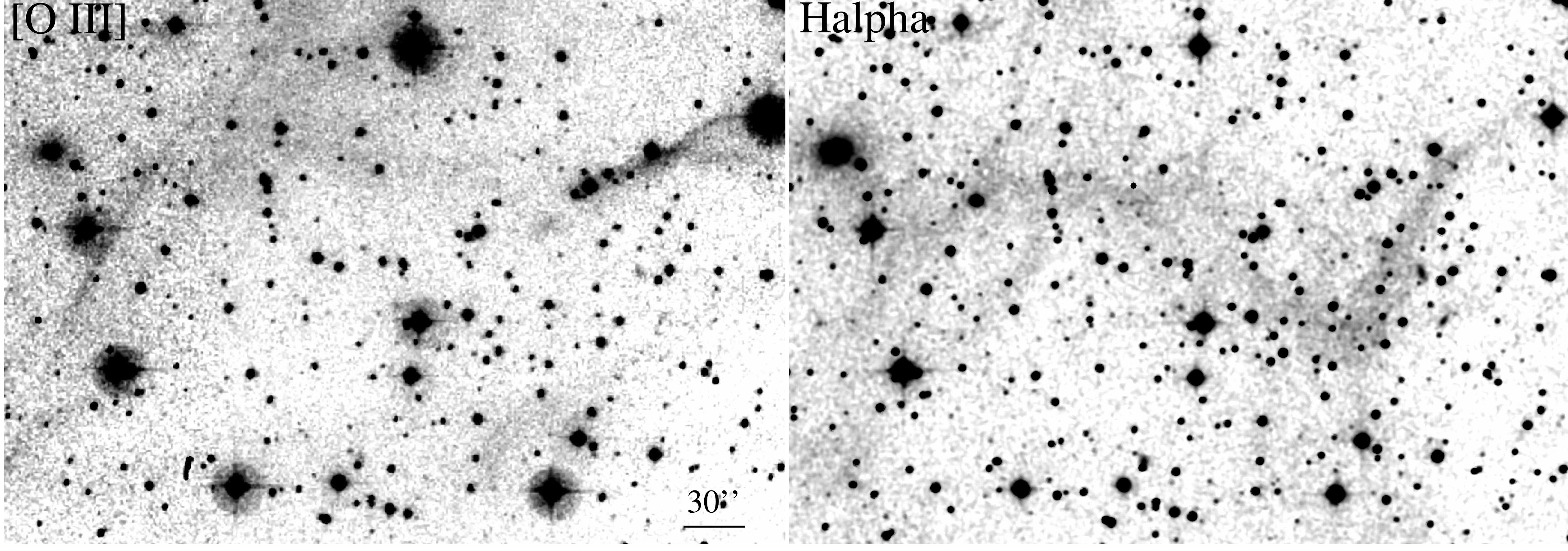}
\caption{Comparison of G179.0+2.6 emissions in [\ion{O}{3}] and H$\alpha$.
{\bf{Top Panels:}}  A 11 hour long exposure of the remnant in  [\ion{O}{3}] (left)
and a 5.8 hour H$\alpha$ image (right) as detected in the 0.4m Orange Observatory telescope. 
North is up, and East is to the left.
The majority of the remnant's optical emission is dominated by strong [\ion{O}{3}] emission
with only a small northeastern filamentary region bright in H$\alpha$.
(Note: Extreme east and west central edges of both images are
contaminated by light leaks.)
{\bf{Bottom Panels:}} Comparison of the remnant's northeastern limb region
as seen in a MDM 1.3m [\ion{O}{3}] image and in the IPHAS H$\alpha$ image.
}
\label{Nic_images}
\bigskip
\end{figure*}

Guided by these wide-field images of the remnant's [\ion{O}{3}] emission,
additional images of the remnant's northeastern and southwestern limbs were
obtained in Feburary and March 2017 using the 1.3m McGraw-Hill telescope at MDM
Observatory at Kitt Peak, AZ.  These were taken using a [\ion{O}{3}] filter
(FWHM $=50$ \AA) using either a 2k $\times$ 2k or a 1k $\times$ 1k SITe
CCDs. On-chip binning yielded an image scale of 1.06 arcsec pixel$^{-1}$. The
image of the SW region was obtained from 4 $\times$ 1500 s exposures, while two overlapping NE
regions were obtained using 2 $\times$ 1200 s exposures and taken under thin cloud conditions.

Low-dispersion spectra of four filamentary regions in the NE and SW were
obtained in January and February 2017 with the 2.4m Hiltner telescope at MDM Observatory  and
the OSMOS Spectrograph \citep{Martini2011}. Using a blue VPH grism (R = 1600),
single 2000 s exposures were taken of four filamentary regions covering 3900--6800
\AA \  with a spectral resolution of 1.2 \AA \ pixel$^{-1}$. 

Standard pipeline data reduction of MDM images and spectra using IRAF was
preformed.  Direct images were bias-subtracted, flat-field corrected using
twilight sky flats, and averaged to remove cosmic rays and improve
signal-to-noise. Spectra were similarly reduced using IRAF and the software
L.A. Cosmic \citep{vanDokkum2001} to remove cosmic rays.  Spectra were
calibrated using an Ar lamp and spectroscopic standard stars
\citep{Oke74,Massey90}. 

\begin{figure*}[t]
\centering
\includegraphics[scale=0.91]{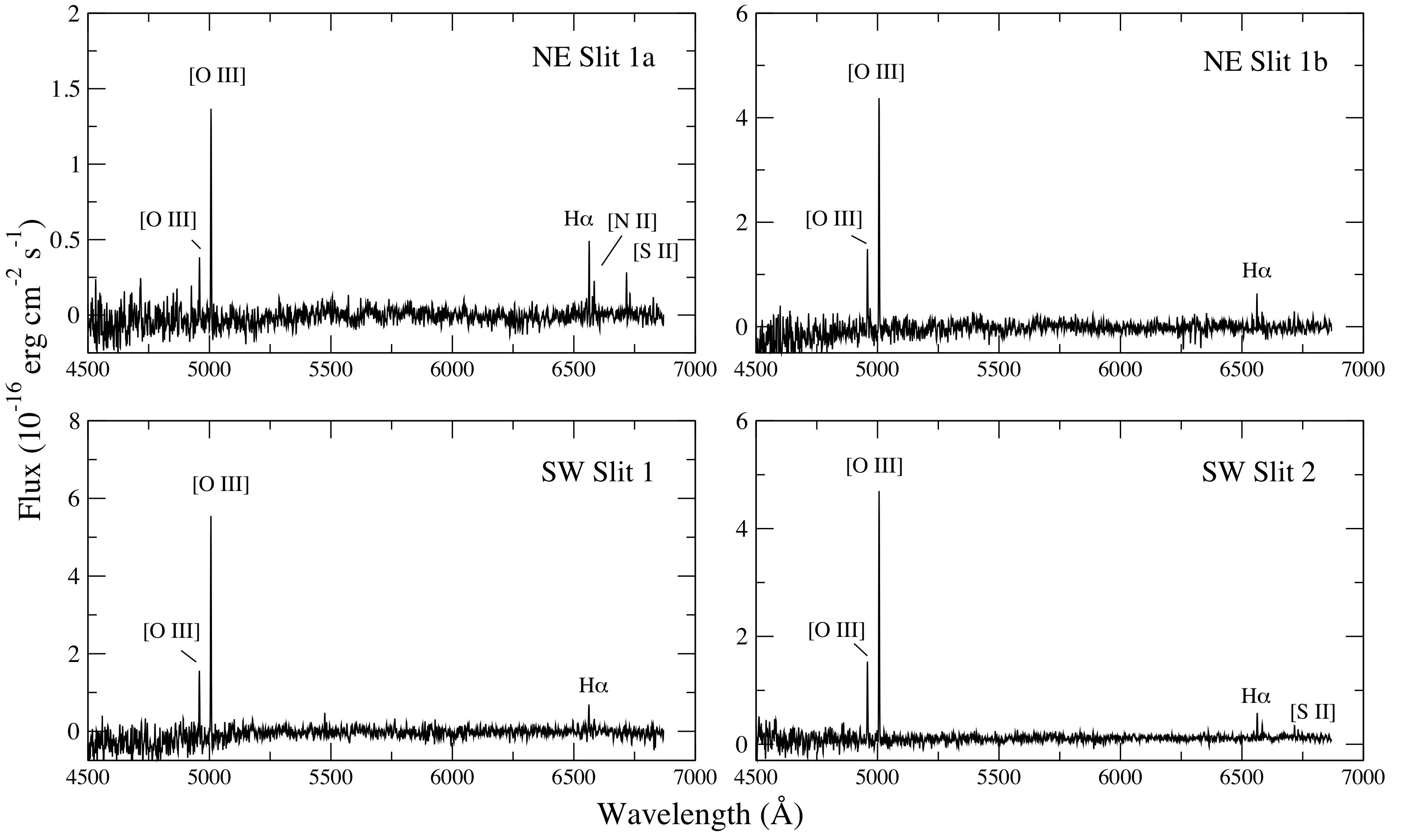}
\caption{Spectra of four regions in G179.0+2.6. See Figure 3 for locations of slit positions. }
\label{spectra}
\end{figure*}

\bigskip

\section{Results}

\subsection{Images} 

Figure~2 shows G179's [\ion{O}{3}] emission structure
along with an overlay of the remnant's radio emission contours as seen by the
Effelsberg 100 m radio telescope at 11 cm (2695 MHz) \citep{Fuerst90}. The remnant's optical
emission forms a nearly complete shell, brightest along its northeastern,
southern, and southwestern limbs.  Although most of the [\ion{O}{3}] emission
is diffuse, some sharp filaments are present along the northeast and southwest limb
regions as shown in Figure~3.

Figure~\ref{Nic_images} shows a comparison of the [\ion{O}{3}] and H$\alpha$ images of G179.
These images make it clear that the remnant's
optical emission is dominated by [\ion{O}{3}] emission.  
Except for the northeastern limb, little H$\alpha$ emission is detected along
most of the remnant's radio shell.  This helps explain why the remnant was
missed in previous H$\alpha$ surveys of Galactic SNRs searching for associated
optical emission.  Not visible in the [\ion{O}{3}] images,  some
exceedingly faint and diffuse H$\alpha$ emission appears present inside the
remnant's lower weatern interior region.

In the few regions where some H$\alpha$ emission is detected, its morphology is
different from that of the remnant's [\ion{O}{3}] emission.  For example, as shown
in the lower panels of Figure~\ref{Nic_images} for the remnant's northeastern
limb area, one finds little correspondence between filaments bright in
[\ion{O}{3}] to those bright in H$\alpha$. The broad and extended ``V''
shaped H$\alpha$ emission feature seen here from the IPHAS image survey (lower
right panel) exhibits little coincident [\ion{O}{3}] emission. Conversely, the
bright [\ion{O}{3}] filament some 30$^{\prime\prime}$ to the north from this
H$\alpha$ filament is largely absent in H$\alpha$ emission.  Weak H$\alpha$
line emission is a property present throughout the remnant's
extensive [\ion{O}{3}] emission structure. 

\subsection{Optical Spectra}

Due to the remnant's faint optical emissions, low-dispersion spectra were only
obtained of the remnant's brighter northeastern and southwestern filaments.
Slit positions are indicated in Figure~3, with the resulting spectra
presented in Figure~\ref{spectra}. Observed emission line fluxes relative to
H$\alpha$ are listed in Table 1.

The [\ion{O}{3}] and H$\alpha$ images of G179 had already suggested unusually
high [\ion{O}{3}]/H$\alpha$ ratios and this was confirmed by spectra. We
measured values around 10 for three of the four regions sampled (see Table 1).
The lack of detected H$\beta$ 4861 \AA \ emission prevents tabulating the
strength of line emissions relative to H$\beta$, the usual ratio used to
express the relative strength of [\ion{O}{3}] emission in SNRs.

In places where the optical emission was bright enough to detect [\ion{S}{2}]
line emission, we find  [\ion{S}{2}]/H$\alpha$ ratios
$>$0.4, i.e., above the standard criteria for identifying shock emission
\citep{Dodorico80,Blair81,Dopita84,Fesen85}.  However, the remnant's faintness
prevented a firm measurement of the ratio of the density sensitive [\ion{S}{2}]
6716, 6731 \AA \ emission lines.  For one slit position, namely NE~1a, we measured a
6716/6731 ratio $\simeq$ 1.30 which indicates a density close to the low
density limit, $\leq$ 100 cm$^{-3}$ \citep{Oster06}.  

\bigskip

\section{Discussion}

With an angular diameter of $\simeq$ 70$^{\prime}$, G179 ranks among the
largest Galactic SNRs known. Moreover, at a distance of $\sim$ 3.5 kpc
\citep{Fuerst86}, this translates into a linear diameter $\approx$70 pc, indicative of  
a relatively old SNR and one likely in the later Sedov-Taylor phase of
adiabatic expansion. The presence of an unusually slowly
rotating gamma-ray pulsar, PSR J0554+3107, with a characteristic spin down age
around 52 kyr \citep{Pletsch13} is consistent with the picture of an well
evolved SNR.

On the other hand, optical emission that is so dominated by 
[\ion{O}{3}] line emission is relatively rare in Galactic remnants, especially large and
old ones like G179. This is because [\ion{O}{3}]/H$\alpha$ ratios around 2 or larger are generated by
relatively fast shocks above $\approx$100 km s$^{-1}$ \citep{Raymond79,Shull79,Hart87}, 
unusual in large and highly evolved SNRs.
Moreover, for there to be weak associated H$\alpha$ emission, the shock's
recombination zone must be `incomplete', meaning the entire postshock cooling
zone is not fully established. Such incomplete shocks are often identified by
[\ion{O}{3}]/H$\beta$ ratios greater than 10 and commonly associated with shocks 
in excess of 100 km s$^{-1}$.

In the case of the optical emission seen in G179 where the emission is so
faint, H$\beta$ emission was not readily detected at any of our four slit positions, 
meaning [\ion{O}{3}] 4959+5007/H$\beta$ ratios are likely quite high, exceeding
20 for positions SW~2 and NW~1b. Such strong 
[\ion{O}{3}] emission filaments have been seen in filaments in a few
remnants \citep{Hester94}.  One example lies in the Cygnus Loop where a
filament along the remnant's far eastern boundary shows a [\ion{O}{3}]/H$\beta$
of 45, near the upper end of the range observed in SNRs \citep{Fesen96}. It is believed this
filaments represents material only recently shocked by the remnant's 300+ km
s$^{-1}$ shock front and has not yet had time to establish a full recombination zone.

However, examples of SNRs whose whole optical emission is dominated by [\ion{O}{3}] line emission are rare.
One large and old galactic SNR whose optical emission is like that of
G179, is G65.3+5.7. It exhibits strong [\ion{O}{3}] emission in almost all of its optical
structure.  This remnant was only discovered through deep 
[\ion{O}{3}] images \citep{Gull77}, and subsequently confirmed by radio
observations \citep{Reich79}. At an estimaged distance of 0.8 -- 1.0 kpc \citep{Gull77,Reich79,Rosado81,Sitnik83}, 
its 3.3 $\times$ 4 degree angular size translates to a physical size $\sim$ 70 pc, similar to G179. 

\begin{deluxetable}{lrrrr}
\tablecolumns{5}
\tablecaption{Relative Fluxes in G179.0+2.6}
\tablehead{
\colhead{Line/Ratio} & \colhead{SW 1} & \colhead{SW 2} & \colhead{NW 1a} & \colhead{NW 1b} }
\startdata
~ H$\beta$ 4861      & $<100$  & $<50$  & $<20$  &  $<50$  \\
~  [O III] 4949      &  280  &  380  & 110  &  330    \\
~  [O III] 5007      &  910  & 1150  & 390  &  1010    \\
~ H$\alpha$ 6563     & 100   & 100   & 100  &  100  \\
~ [N II] 6583        & 40    &  52   &  40  &  33  \\
~ [S II] 6716        & $<25$ & 44    & 56   &  $<30$  \\
~ [S II] 6731        & $<25$ & 34    & 25   &  $<30$   \\
~ [S II]/H$\alpha$   & \nodata   & 0.78 & (0.80) & \nodata     \\
~ [S II] 6716/6731   & \nodata   & 1.30 & $>1.3$ & \nodata     \\
~ $\rho$ (cm$^{-3}$) & \nodata   &$<$100&$<$100 &  \nodata     \\
~ H$\alpha$ flux\tablenotemark{a} & 2.1  & 1.4 & 1.3  & 1.5   \\
\enddata
\tablenotetext{a}{Flux units: $10^{-16}$ erg s$^{-1}$ cm$^{-2}$. }
\end{deluxetable}

Except for a small filamentary region known as Sharpless 91 (S91;
\citealt{Sharpless59}) along the southern limb of its $4 \times 3.3 $ degree
shell,  most of G65.3+5.7's filaments show large
[\ion{O}{3}]/H$\beta$ ratios consistent with incomplete recombination zones and
shock velocities above 90 km s$^{-1}$ \citep{Fesen85,Mav02}. The one noticeable
exception is the H$\alpha$ bright filament, S91. This situation is similar to what is seen
in G179 whose only bright H$\alpha$ emission is a few filaments along the remnant's
northeastern limb (see Fig.\ 4). 

Based on [\ion{O}{3}] line profiles, \citet{Boumis04} estimated a global
expansion velocity for G65.3+5.7 of $\approx$155 km s$^{-1}$. This value is
consistent with a relatively high-shock velocity through its structure despite
its old age. A similar analysis and shock velocity probably applies to
G179.

For both G65.3+5.7 and G179, a relatively low surrounding  ambient
density is likely the cause for their optical emission to be nearly dominated
by strong [\ion{O}{3}] filaments, an otherwise rare property in older SNRs.  A
low density surrounding ISM, as has been suggested in the case of G65.2+5.7
caused by a pre-SN progenitor wind-blown cavity \citep{Xiao09}, would also help
explain why such [\ion{O}{3}] emission dominate remnants are also so optically
faint.

The distance to G179 is uncertain.  A value of 3.5 kpc was cited by
\citet{Fuerst86} based on the $\Sigma-D$ relation given by \citet{Milne79}.
However,  $\Sigma-D$ derived distances are controversial and
widely viewed as not being especially accurate \citep{Green91,Dubner15}. Indeed, a search of
recent $\Sigma-D$ distance estimates for G179 finds a wide range of estimated distance
values; \citet{Case98} lists 6.1 kpc, \citet{Gus03} cite a value of 2.9 kpc, while
\citet{Pavlovic14} lists a distance of 3.1 kpc but with a large
uncertainty range of 2.1--8.2 kpc.

Although its true distance is currently uncertain, G179 is unlikely to lie at distances
much greater than $\sim$ 5 kpc. The remnant's [\ion{O}{3}] emission is so
strong relative to H$\alpha$, foreground extinction toward the remnant is
likely to be modest, a conclusion consistent with its anti-center galactic
location. Reddening at $l = 179.0 \pm 0.3, b = 2.6 \pm 0.3$ based on a
three-dimensional map presented in \citet{Green2015}, which utilizes Pan-STARRS
\citep{Schlafly2014} and 2MASS \citep{Skrutskie2006} photometry, indicate G179
lies in a region of relatively low extinction with $E(B-V)$ values around 0.2
at 1 kpc rising above 0.4 around 5 kpc. Values of  $E(B-V)$ much greater than
0.3 would raise the [\ion{O}{3}]/H$\alpha$ above that seen in even extreme cases
of incomplete recombination.

In addition, at a distance of 5 kpc the remnant's diameter would exceed 100 pc
making it one the physically largest known Galactic SNRs. Such a size would be hard to
understand given the requirement of a shock velocity well in excess of 100 km
s$^{-1}$ enabling it to exhibit incomplete recombination in the postshock
cooling zone. 

\section{Conclusions}

We have discovered substantial optical emission from the large Galactic
supernova remnant G179.0+2.6 which previously had no known associated optical
emission.  Narrow passband filter images reveal a nearly complete but faint
emission shell dominated by strong [\ion{O}{3}] line emission.  Equally deep
H$\alpha$ images detect little coincidence emission indicating an unusually widespread
high [\ion{O}{3}]/H$\alpha$ ratio for such a large and apparently old remnant.
Low-dispersion optical spectra of several regions confirm large
[\ion{O}{3}]/H$\alpha$ line ratios, suggesting incomplete cooling in the
postshock recombination zone.  The remnant's unusually strong [\ion{O}{3}]
emission suggests shock velocities above 100 km s$^{-1}$ throughout most of the
remnant, likely reflecting a relatively low density ambient ISM.

\acknowledgements

We thank Sakib Rasool for helpful communications regarding 
optical emission associated with G179, and the MDM Observatory
staff for their excellent instrument assistance.  This research was made
possible by funds from the NASA Space Grant, the Jonathan Weed Fund, the Denis
G. Sullivan Fund, and Dartmouth's School of Graduate and Advance Studies.

\end{document}